\documentstyle[11pt,epsfig,paspconf]{article}

\newcommand{\bi}{\begin{itemize}\tighten}
\newcommand{\ei}{\end{itemize}}

\newcommand{\be}{\begin{equation}}
\newcommand{\ee}{\end{equation}}
\newcommand{\bea}{\begin{eqnarray}}
\newcommand{\eea}{\end{eqnarray}}

\newcommand{\nn}{\nonumber}

\newcommand{\Bphis}{B_{\phi S}}
\newcommand{\Bzs}{B_{z S}}
\newcommand{\Ps}{P_S}

\newcommand{\Rs}{{R_S}}

\newcommand{\Pave}{\langle P \rangle}
\newcommand{\sigsqave}{\langle \sigma^2 \rangle}

\newcommand{\W}{{\cal W}} 
\newcommand{\M}{{\cal M}}
\newcommand{\V}{{\cal V}}

\newcommand{\mvir}{m_{vir}}  

\begin{document}

\title{A New Model for Filamentary Molecular Clouds}
\author{Jason D. Fiege and Ralph E. Pudritz}
\affil{Dept. of Physics and Astronomy, McMaster University, Hamilton, ON}

\begin{abstract}
We develop a theory for filamentary molecular clouds including the effects of ordered magnetic fields,
and external pressure.  We first derive a new {\em virial equation} appropriate for filamentary
clouds.  By comparing with observational results collected from the literature, we find that the fields are likely {\em helical}.  Secondly, we construct
numerical, MHD models of filamentary clouds that agree with the observational constraints.  
We find that our models
produce more realistic density profiles $r\sim r^{-1.8~to~-2}$ than previous models,
where the density falls off as $r^{-4}$.
\end{abstract}

\section{Introduction}
\label{sec:intro}
Most molecular clouds are {\em filamentary} structures that are supported by
non-thermal, MHD turbulence, as
well as large scale ordered magnetic fields (eg. review, McKee et. al 1993).
Nevertheless, virtually all theoretical models assume spheroidal geometry.  While spheroidal models
obviously apply to molecular cloud {\em cores}, they cannot adequately describe molecular clouds 
on larger scales.  

Observations suggest that
some filamentary clouds may be wrapped by helical fields (Heiles 1987; Bally 1987).
A few authors have previously modeled filamentary clouds with helical fields 
(cf. Nakamura, Hanawa, \& Nakano 1993; Hanawa et al., 1993), but the fields in these
models simply rescale the Ostriker (1964) solution for unmagnetized filaments.
They are also unconstrained by observational data.

The role of the external pressure has been almost completely ignored by previous models of filamentary
clouds.  As we discuss in Section \ref{sec:Pressure}, real molecular clouds are truncated at finite radius 
by the pressure of the external medium.  By ignoring the external pressure, most existing models cannot 
adequately describe real molecular clouds.

\section{Surface Pressures on Molecular Clouds}
\label{sec:Pressure}
Molecular clouds and the surrounding atomic gas are dominated by non-thermal motions, which result in
{\em total} pressures that greatly exceed the thermal pressure. The total pressure of the atomic gas has been evaluated at the 
Galactic midplane by Boulares and Cox (1990),
who find pressures on the order of $10^4~K~cm^{-3}$.  However, it is likely that the molecular clouds are surrounded
by atomic gas at significantly higher pressures.  Some are associated with HI clouds at a pressure of 
$\sim 10^5~K~cm^{-3}$ (Chromey, Elmegreen, and Elmegreen 1989).  
Therefore, we conservatively adopt surface pressures in the range of $10^{4-5}~K~cm^{-3}$ for our analysis.  

\section{A New Form of the Virial Equation for Filamentary Molecular Clouds} 
\label{sec:vir}
We consider the virial equilibrium of a long filamentary cloud with an
external pressure $\Ps$ that {\em truncates} the cloud at some cylindrical radius $\Rs$, and a 
general magnetic field of helical geometry.
By considering only radial equilibrium, we derive a new virial equation from the 
{\em tensor} virial theorem:
\be
\frac{\Ps}{\Pave}=1-\frac{m}{\mvir}\left(1-\frac{\M}{|\W|}\right),
\label{eq:virial}
\ee
where
\bea 
\Pave &=& \frac{\int P d\V}{\V} \nn\\
\mvir &=& \frac{2\sigsqave}{G} \nn\\
\W &=& = -m^2 G \nn\\
\M &=& \frac{1}{4\pi}\int B_z^2 d\V - \left(\frac{\Bzs^2+\Bphis^2}{4\pi} \right) \V .
\eea
In this equation, $\Ps$ and $\Pave$ are the surface and average internal
pressures, $m$ is the mass per unit lengh, $\mvir$ is the ``virial'' mass per unit length.
$\W$ and $\M$ are respectively the gravitational and {\em total} magnetic energies per unit length.
We note that $\W$ is independent of the filament radius.  

The total magnetic energy $\M$ may be either positive or negative
depending on whether the poloidal field or the toroidal field dominates the overall energetics of
the cloud.  Thus, the poloidal field component helps to {\em support} the cloud radially against self-gravity, while
the toroidal component works with gravity and the external pressure to squeeze the cloud.

The virial quantities $m/\mvir$ and $\Ps/\Pave$, in equation \ref{eq:virial}, may be 
determined from observations; thus, we may easily {\em infer} the magnetic parameter $\M/|\W|$.  
We have calculated $m/\mvir$ and $\Ps/\Pave$ for several filaments, based on observational results gathered from the literature.
The reader is referred to Fiege \& Pudritz (1999a) for the full data tables and references.
We refer to Figure 1 of Pudritz and Fiege 
(1999), in this volume, where we plot contours of $\M/|\W|$ over these virial parameters.
We find that most of the filaments in our sample fall in the range
\bea
0.11 &\le& m/\mvir \le 0.43 \nn\\
0.012 &\le& \Ps/\Pave \le 0.75,
\label{eq:obs}
\eea
and that $\M/|\W|<0$, which is consistent
with a helical magnetic field.

\section{Numerical Magnetostatic Models}
\label{sec:helix}
Our theoretical models involve three parameters, two to describe the mass loading of the
poloidal and toroidal field lines, and a third that specifies the radial concentration of the filament. 
We fully sample our parameter space by a Monte Carlo method, in which models are randomly generated
and tested for agreement with the observational constraints given
in equation \ref{eq:obs}.  
In addition to
equation \ref{eq:obs}, we demand that the magnetic and kinetic energies are nearly in equipartition;
$ 0.2 \le \frac{M}{K} \le 5$,
where $M$ and $K$ are the average magnetic and kinetic energy densities in the cloud.
This has been observationally determined for many clouds (Myers \& Goodman
1988a,b).

\begin{figure}
\begin{minipage}{.46\linewidth}
\epsfig{file=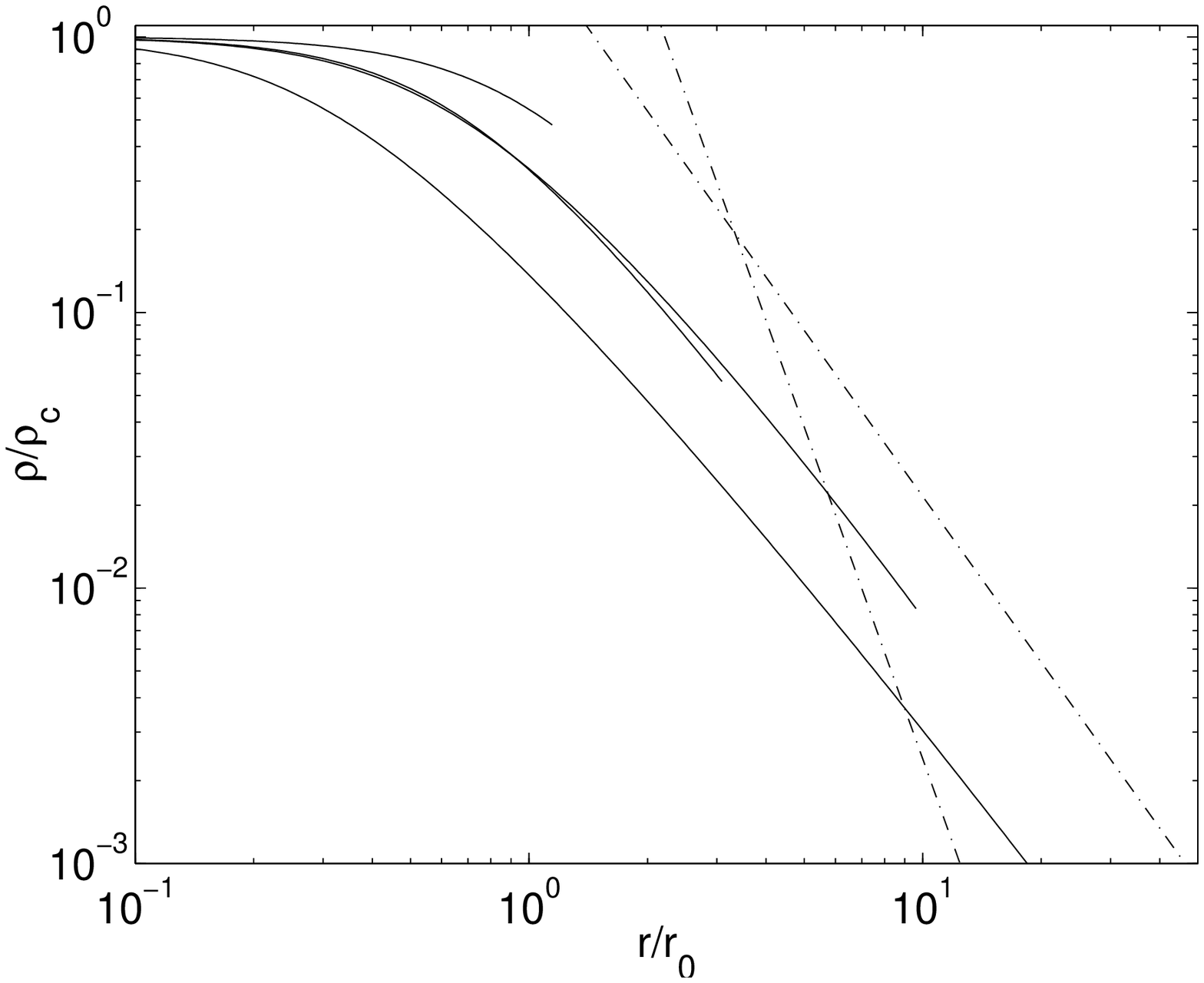,width=\linewidth}
\end{minipage}
\begin{minipage}{.53\linewidth}
\epsfig{file=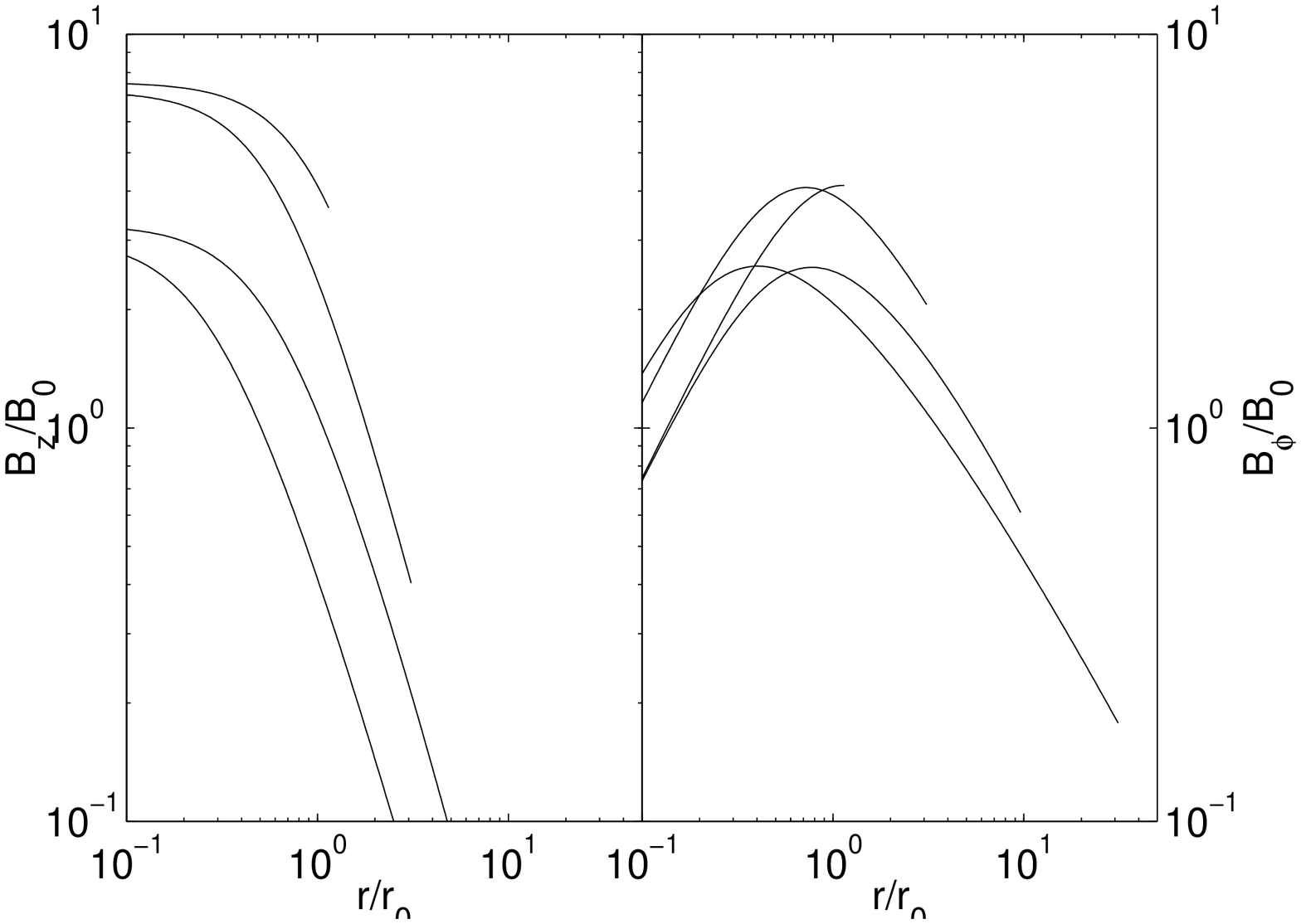,width=\linewidth}
\end{minipage}
\caption{The density and magnetic structure of our models.  
The dashed lines represent $r^{-4}$ and $r^{-2}$ profiles for comparison.}
\label{fig:structure}
\end{figure}

Figure \ref{fig:structure} shows the structure of a few models that are consistent with our
constraints.  We find that the density falls off as $r^{-1.8}$ to
$r^{-2}$ in the outer regions, which is in excellent agreement with recent results by
Alves et al. (1998) and Lada et al. (1998) for the L977 and IC5146 filamentary molecular clouds.
We find that the peak poloidal field in our models is always much stronger than the peak toroidal
field; only the outer regions
are dominated by the toroidal field component.  
In this sense, our helical fields are actually quite weakly wrapped.  These results are
robust; a Monte Carlo sampling of our parameter space shows that nearly {\em all} allowed models 
have these characteristics.

\section{Stability of Filamentary Molecular Clouds}
We consider the stability of filamentary molecular clouds against axisymmetric fragmentation into cores.  
By solving the linearized equations of MHD and self-gravity, we determine how pressure truncation 
and the individual magnetic field components affect the stability of our models.
We find that pressure truncation has a stabilizing effect, which results from the decreased mass
per unit length of the filament.
Both field components also stabilize clouds against gravitational fragmentation.  However, we find
that sufficiently strong toroidal fields may trigger MHD-driven instabilities.
We refer the reader to Fiege \& Pudritz (1999b), where we fully explore the stability of our models.

\section{Discussion}
We have found indirect evidence, based on a very general virial analysis, suggesting that most
filamentary molecular clouds may be wrapped by helical magnetic fields.  We have also constructed
numerical MHD models that agree with the observations.  These models differ from previous models in that they 
have much shallower density profiles, which are in good agreement with the available observational results;
we find that the density falls off as $r^{-1.8}$ to $r^{-2}$, compared with the $r^{-4}$ behaviour of
previous models.  This behaviour is entirely due to the toroidal character of the magnetic field 
in the outer regions.  We also find that models with purely poloidal fields have steep 
density gradients that are not allowed by the observations.


\begin{references}
\reference Alves J., Lada C.J., Lada E.A., Kenyon S.J., Phelps R., 1998, Ap.J., 506, 292
\reference Bally J., 1989, in Proceedings of the ESO Workshop on Low Mass Star Formation and Pre-main Sequence Objects,
        ed. Bo Reipurth (Garching:European Southern Observatory), p.1
\reference Boulares A., Cox D.P., 1990, \apj, 365, 544
\reference Chromey F.R., Elmegreen, B.G., Elmegreen, D.M., 1989, \apj, 98, 2203
\reference Fiege, J.D., \& Pudritz, R.E., 1999a, \mnras, astro-ph/9901096
\reference Fiege, J.D., \& Pudritz, R.E., 1999b, \mnras, astro-ph/9902385 
\reference Hanawa T., et al., 1993, \apj, 404, L83
\reference Heiles C., 1987, Ap.J., 315, 555
\reference Lada C.J., Alves J., Lada E.A., 1998, Ap.J. in Press
\reference McKee C.F., Zweibel E.G., Goodman A.A., Heiles C., 1993,
	in Protostars and Planets III, ed. Levy E.H. \& Lunine J.I. (Tucson:University of Arizona Press), 327
\reference Myers P.C., Goodman A.A., 1988a, \apj, 326, L27
\reference Myers P.C., Goodman A.A., 1988b, \apj, 329, 392
\reference Nakamura F., Hanawa T., Nakano T., 1993, \pasj, 45, 551
\reference Ostriker J., 1964, \apj, 140, 1056
\reference Pudritz R.E., Fiege J.D., 1999, in Proceedings of the Naramata Workshop on the Interstellar Medium
\end{references}
\end{document}